# First experiment on fission transients in highly fissile spherical nuclei produced by fragmentation of radioactive beams

C. Schmitt, [1,†] P.N. Nadtochy, [2] A. Heinz, [3] B. Jurado, [4] A. Kelić, [1] K.-H. Schmidt [1]

[1] *GSI, Planckstraße 1, 64291 Darmstadt, Germany*
[2] *Omsk State University, Department of Theoretical Physics, Mira Prospect 55-A, 644077 Omsk, Russia*
[3] *Wright Nuclear Structure Laboratory, Yale University, 272 Whitney Avenue, New Haven, CT 06520, United States*
[4] *Université Bordeaux I, CNRS/IN2P3, CENBG, Chemin du Solarium, BP 120, 33175 Gradignan, France*

We report on a novel experimental approach for studying the dissipative spreading of collective motion in a meta-stable nuclear system, using, for the first time, highly fissile nuclei with spherical shape. This was achieved by fragmentation of 45 radioactive heavy-ion beams at GSI, Darmstadt. The use of inverse kinematics and a dedicated experimental set-up allowed for the identification in atomic number of both fission fragments. From the width of their charge distributions, a transient time of $(3.3 \pm 0.7) \cdot 10^{-21}$ s is deduced for initially spherical nuclei.

The problem of escape from a meta-stable state appears in many fields as various as fluid mechanics, chemistry or nuclear physics [1]. An excellent test case for such a process is nuclear fission. Once an excited system has been produced, the collective coordinates start adjusting to the potential-energy landscape, and the distribution of shapes inside the fission saddle develops towards thermo-dynamical quasi-equilibrium. After some time delay, all states in the quasi-bound region are populated according to the available phase space, including those above the fission saddle. Once the system leaves the quasi-bound region, it is driven by the dominating Coulomb force to more elongated shapes and finally separates into two fragments.

The ideal scenario for studying the dynamics of the fission process is described in the pioneering theoretical work of Grangé, Weidenmüller and collaborators [2, 3]: They have chosen an initial excited system characterized by a spherical shape. In this case, the probability distribution starts from a configuration where the level density has a local maximum. Under this specific initial condition, the maximum of the probability distribution does not move. The distribution only spreads out under the influence of dissipation due to the *fluctuating forces*. In contrast, the evolution from saddle to scission is characterized by a strong driving force, and it is the *friction force* which dominates the dissipation mechanism by slowing down the directed motion towards scission and increasing the saddle-to-scission time relative to its non-viscous limit. Thus, in this ideal scenario starting from the bottom of a spherical potential well, the two stages of the fission process probe selectively the two dissipative phenomena, diffusion and friction, in nuclei [4]. The present work is motivated by the idea to study the time scale of nuclear diffusion.

---

[†] Present address : Université Lyon I, CNRS/IN2P3, IPNL, Rue Enrico Fermi, 69622 Villeurbanne, France.

Attempts to selectively investigate the purely dissipative relaxation process inside the quasi-bound region according to the aforementioned ideal scenario were hampered by the unavailability of spherical highly fissile nuclei. In the present work, we report on a novel experimental approach, which comes very close to these conditions, by investing a considerable instrumental effort, utilising the experimental installations available at GSI, Darmstadt. This approach is based on a two-step reaction mechanism illustrated in the upper part of Fig. 1. Fragmentation of a primary $^{238}$U beam at 1 $A$ GeV in a beryllium target produces a large variety of nuclei. Most of these products adapt to their ground-state configuration well before reaching the second target. Among them, radioactive isotopes of elements between astatine and thorium, with neutron numbers around the $N = 126$ shell, are particularly interesting, because they are spherical in their ground state. These nuclei, still at relativistic energies, act as secondary projectiles, which further impinge on a lead target. There, fragmentation of the radioactive beams yields fissile pre-fragments which reach up to high excitation energies [5, 6] and are characterized by small angular momenta [7]. Within the fast cut-off picture [5], we estimated the quadrupole deformation parameter $β_2$ induced by the abrasion process. Fig. 2 shows this quantity as a function of the pre-fragment mass $A_{prf}$. The initial $β_2$ of the nuclei which end in fission, and whose calculated abundances are displayed in the inset of the figure, remains small and oblate ($|β_2|<0.15$). The excited pre-fragment finally decays by a competition between fission and particle evaporation.

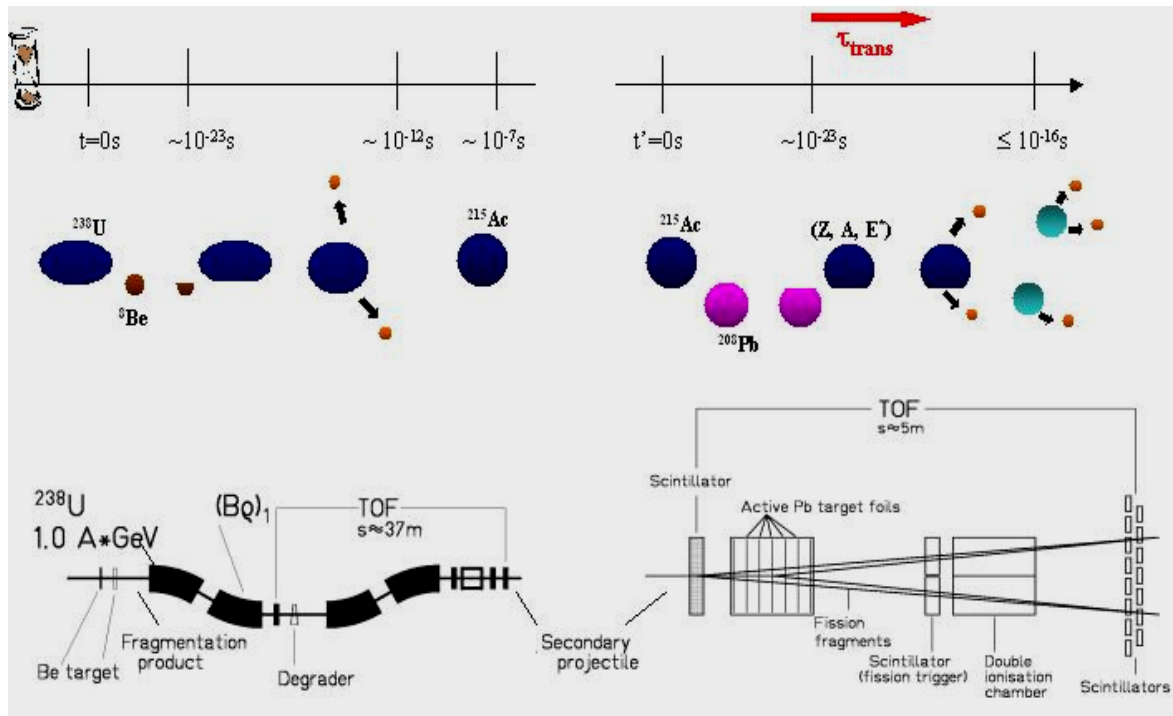

**Figure 1:** Top: Illustrative drawing of the radioactive-beam production (left) and the subsequent fragmentation-fission reaction (right). Typical time scales t and t' referring to the primary and secondary reaction, respectively, are indicated. Bottom: Experimental set-up involved along the outlined reaction scenario, consisting of the FRS (left) and the fission set-up (right).

A schematic drawing of the experimental set-up is given in the lower part of Fig. 1. In a first stage, the fragment separator (FRS) located behind the primary target, with its ability to spatially separate and identify ions event by event, was used to prepare secondary beams from the $^{238}$U fragmentation products. At the exit of the FRS, around 60 proton-rich projectiles ranging from $^{205}$At to $^{234}$U at about 420 $A$ MeV were available. We focus here on 45 of these radioactive beams, which are spherical in their ground state. As shown in Fig. 2, the nuclei produced in the second stage of the set-up by peripheral collisions of such projectiles with lead target nuclei are still nearly spherical. If their de-excitation ends with fission, the inverse kinematics of the reaction allows an accurate determination of the atomic numbers of both fission fragments *simultaneously* in a double ionisation chamber with a resolution of $\Delta Z_{1,2} = 0.4$ (FWHM). The excellent resolution is clearly visible in the ($Z_1$, $Z_2$) charge-correlation spectrum presented in Fig. 3 – each spot corresponds to one pair of fragments. We refer to [8] for a detailed description of the set-up.

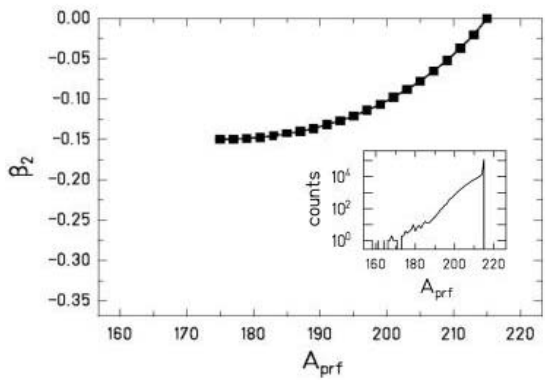 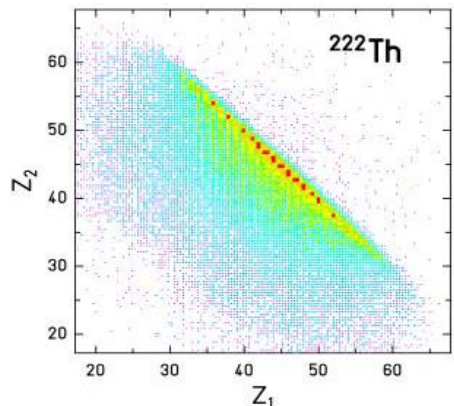

**Figure 2**: Calculated quadrupole deformation parameter $\beta_2$ of pre-fragments of mass $A_{prf}$ resulting from abrasion of a spherical beam with mass $A_{proj}$ = 215. The inset shows the calculated mass distribution of all pre-fragments originating from the abrasion of $^{215}$Ac in Pb and which end up in fission. The peak at $A_{prf}$ = 215 is due to electromagnetic interactions, see [A. Heinz et al., Nucl. Phys. **A 713**, 3 (2003)].

**Figure 3**: Experimental correlation between the charges ($Z_1$, $Z_2$) of the two fission fragments for a $^{222}$Th beam. Note the logarithmic z-scale.

Fragmentation of heavy projectiles leads to a large variety of pre-fragments which can be classified using pertinent filters. Here, we profit from the measurement of the sum $Z_1+Z_2$ of the charges of the two fission fragments. Full-acceptance experiments [9] have shown that in peripheral collisions at relativistic energies the impact parameter and the nuclear charge of the heaviest fragment measured in a given event are strongly correlated. This also implies a close relation between the charge of the fissioning nucleus $Z_{fiss}$ and the impact parameter. Because the probability for the emission of light charged particles between saddle and scission as well as after scission is small for the systems considered in this work, the $Z_1+Z_2$ sum is equal to the charge $Z_{fiss}$ of the fissioning nucleus in most cases. Hence, $Z_1+Z_2$ is correlated with the impact parameter and it can be used as a measure of the excitation energy induced in the reaction – a lower $Z_1+Z_2$ corresponds to a smaller impact parameter and, thus, to higher initial excitation energy. The latter increases by about 65 MeV per charge unit lost with respect to the projectile charge $Z_{proj}$ in peripheral collisions [6] before saturating for more central collisions due to the onset of multi-fragmentation [10]. On the average, over the whole set of beams, the initial temperature reaches up to ~5.5 MeV when $Z_1+Z_2$ decreases down to 70 [11]. Furthermore, $Z_1+Z_2$ scales with the fissility $x$ of the excited system, ranging from 0.78 down to 0.66 with decreasing $Z_1+Z_2$.

Following Ref. [12], we use the width of the fission-fragment charge distribution $\sigma_Z$ for establishing a chronometer up to the fission saddle. From a comprehensive compilation of data, the authors of [13] have shown that the width $\sigma_A$ of the fission-fragment mass distribution, and accordingly $\sigma_Z$, scales with the temperature at saddle. The latter is governed by the number of particles de-exciting the system before it reaches the saddle point. Extrapolating the systematic of [13] to higher energies, we demonstrated [12] that $\sigma_Z$ is closely linked to the transient delay $\tau_{trans}$ [2] and, hence, stands for a pertinent clock for the time the system needs to cross the fission barrier. The experimental width $\sigma_Z$ of the fission-fragment charge distribution is shown in Fig. 4 as function of $Z_1+Z_2$ for a sample of secondary spherical projectiles.

For a quantitative analysis of the charge width, we compare the data with results from the Monte Carlo code ABRABLA [5, 6, 14, 15]. This code, which is based on the abrasion-ablation model, is able to simulate the complex reaction scenario realized in the experiment. Its validity has been assessed by investigating various systems and observables [16]. Fission is treated as a dynamical process using a *time-dependent* fission decay-width $\Gamma_f(t)$ that is based on a realistic analytical approximation [17] to the numerical solution of the one-dimensional Fokker-Planck equation for a system initially localized in the minimum of a spherical potential well. Moreover, the ratio $a_f/a_n$ of the level-density parameter at the saddle point to that at the ground state is realistically calculated taking the influence of deformation into account [18, 19]. The experimental dependences of $\sigma_Z$ on $Z_1+Z_2$ for the 45 spherical beams have been compared with calculations performed varying the dissipation strength $\beta$ from $1 \cdot 10^{21} s^{-1}$ up to $7 \cdot 10^{21} s^{-1}$ [20]. As observed in Fig. 4, the whole data set is well described with $\beta = (4.5\pm0.5) \cdot 10^{21} s^{-1}$ independent of $Z_1+Z_2$, i.e. independent of excitation energy and fissility. Also shown in the figure are predictions based on the Bohr-and-Wheeler transition-state model [21] and Kramer's diffusion scenario [22]. While the former completely neglects viscosity, the latter accounts for its influence on the stationary value of $\Gamma_f(t)$ but omits the pre-saddle delay. Both models predict too steep a rise of $\sigma_Z$ with decreasing $Z_1+Z_2$. Due to the absence of transients, there is insufficient time for particle evaporation that would cool the nucleus during $\tau_{trans}$ and, hence, narrow the fission-fragment charge distribution. The overall reduction of the stationary fission-decay width, caused by dissipation and accounted for in Kramer's theory, has a strikingly weaker influence in narrowing the fission-fragment charge distributions than the initial suppression of fission during the transient time has. As soon as $Z_1+Z_2$ differs by more than about 3-4 charge units from the secondary projectile charge $Z_{proj}$, calculations neglecting transients substantially deviate from the data. That sets the lower limit for the sensitivity of $\sigma_Z$ to initial temperatures above about 3 MeV. In addition to the magnitude of $\beta$, it follows for the whole set of data a nearly constant transient time with an average value of $<\tau_{trans}> = (3.3\pm0.7)\cdot10^{-21}$s.

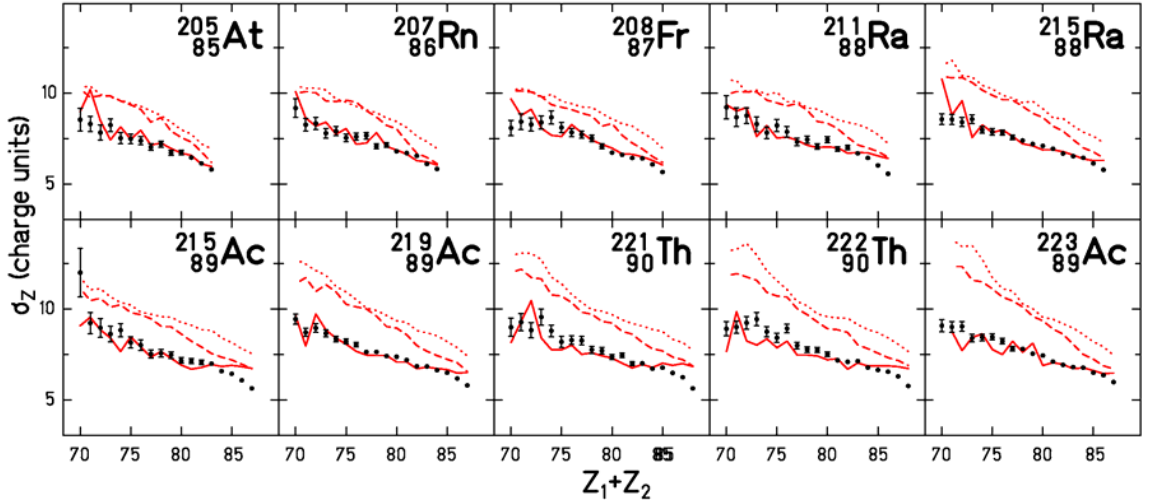

**Figure 4:** Width $\sigma_Z$ as a function of $Z_1+Z_2$ for a sample of spherical secondary beams as indicated. The data (dots) are compared with calculations using Bohr-and-Wheeler- [21] (dotted lines) and Kramers- [22] (dashed lines) fission-decay width as well as the time-dependent approximation (full lines) from [17]. In the two latter cases, $\beta$ is set to $4.5\cdot10^{21}$s$^{-1}$. Staggering in the calculations is due to statistical fluctuations.

In our previous work [12], we investigated dissipative phenomena in the fission of $^{238}$U projectile fragments. For those systems we obtained a transient time of $(1.7\pm0.4)\cdot10^{-21}$s, which is appreciably shorter than the present result. We attribute this discrepancy to the fact that the fissioning nuclei studied in [12] are deformed, since stemming from fragmentation of the primary $^{238}$U projectile, which has a prolate shape characterized by $\beta_2\approx0.23$ [23]. To our knowledge, this constitutes the first *experimental* indication that the time, the system needs to reach the barrier, is shorter when the initial configuration is already closer to the saddle point than when starting from a spherical shape. Our observation is consistent with one-dimensional dynamical calculations [24], which were recently confirmed in a three-dimensional space [25]. They predict that the influence of the initial deformation of the compound nucleus has a strong influence on $\tau_{trans}$. This effect is customarily neglected [24, 26]. The experimental confirmation of such a sizeable influence suggests re-visiting previous analysis, and namely those dealing with heavy-ion fusion-induced fission, where the capture configuration differs significantly from a compact compound nucleus. The value extracted for the dissipation strength might be biased when disregarding this effect.

Spherical initial shapes have also been achieved by spallation reactions for experiments on dissipation in lighter systems, see e.g. [27, 28, 29]. The present work goes a step further by utilising for the first time highly fissile spherical systems for which fission properties can be studied over a much wider range of excitation energy. In addition, for fission of spherical nuclei, it is the only case where the two fragments were identified in atomic number, exploiting the charge width as a clock at the saddle point.

Our result for $\beta$ suggests an over-damped motion at small deformation and high excitation energy, amounting to about 25% of what is predicted by the one-body wall theory [30]. In an extended study of a large body of observables measured in fusion-fission experiments and which to most part are sensitive to the whole path up to scission, dynamical calculations [31] led to $\beta$ values comparable with the present one. More generally, theoretical studies on the surface motion of a cavity wall (see [32] and therein), among which are vibrations and fission, point out the influence of quantal effects and chaoticity in single-particle motion, suggesting that the original derivation of the one-body wall damping mechanism [30] overestimates the dissipation rate in nuclear matter.

We would like adding a word of caution at this point. Systematic investigations [33] have shown that the time scale of the fission process is more realistically modelled by three-dimensional calculations. Therefore, our conclusion drawn on the dissipation strength $\beta$ might be considered as preliminary since it was deduced from calculations which model the time-dependent fission process on the basis of the one-dimensional Fokker-Planck equation. The transient time, however, which is rather directly reflected in our experimental signatures should not be affected.

The present experiment has provided dedicated data on the diffusion inside the quasi-bound region. The consistent description of these data together with the large body of data related to scission times [31] with the same dissipation strength indicates that diffusion and friction inside and outside the quasi-bound region are consistently described by conventional Fokker-Planck or Langevin dynamics with a rather universal dissipation strength. Also the shorter transient time found for initially deformed systems [12] qualitatively fits to this picture. Hence, our data do not evidence any departure from normal diffusion dynamics in nuclear matter. According to Einstein's relation, the nearly constant dissipation strength obtained in this work suggests that the nuclear diffusion coefficient scales with temperature.

In summary, we report on an experiment realized under very favourable conditions for studying transient effects in nuclear fission. For the first time, we succeeded in probing diffusion in *spherical highly excited fissile nuclei*. We benefit from a considerable instrumental effort using projectile-fragmentation in a two-step experiment at GSI, Darmstadt. The analysis of the fission-fragment charge distributions strongly supports the manifestation of pre-saddle transient effects at high excitation energies, which delays the quasi-stationary flow across the fission barrier. A dissipation strength of $\beta = (4.5\pm0.5) \cdot 10^{21} s^{-1}$ has been extracted, corresponding to a transient time of $<\tau_{trans}> = (3.3\pm0.7) \cdot 10^{-21} s$ for an initially spherical configuration. The combination of previous results [31], which are to a great part sensitive to the whole path up to scission, with the present conclusion, restricted to the pre-saddle region, seems to indicate that the dissipation strength does not vary strongly with deformation. Furthermore, no evidence for a noticeable dependence of the dissipation strength on either excitation energy or fissility has been found in the temperature and fissility regime studied here, suggesting a one-body viscous mechanism. To further investigate temperature- and fissility-dependence, a more accurate characterisation of the decaying system is highly desirable. Such studies are foreseen at the future FAIR facility [34] by detecting the light particles emitted in coincidence with fission and identifying the fragments in charge *and* mass.

**Acknowledgements**

We are grateful to P. Armbruster for fruitful discussion. Two of us (C. S. and P.N.N.) are grateful for the post-doctoral position at GSI granted by the A. von Humboldt foundation. This work has been financially supported in part under U.S. DOE Grant Number DE-FG02-91ER-40609 and by the French-German GSI/IN2P3/CEA collaboration under Contract Number 04-48.